\documentclass{aa}     
\usepackage{epsfig}

\def\a4{ABCG~496{}}
\def\bj{$b_{\rm J}${}}
\thesaurus{11.03.4 ABCG~496; 11.03.1 }

\title{A photometric catalogue of galaxies in the cluster Abell 496.
\thanks{Based on plates scanned with the MAMA microdensitometer at CAI, Paris
and on observations collected at the European Southern Observatory, La Silla,
Chile}
\thanks{Tables 1 and 2 are only available in electronic form at the CDS via
anonymous ftp to cdsarc.u-strasbg.fr (130.79.128.5).}
}

 \author {
  E.~Slezak \inst{1}
\and
  F.~Durret \inst{2,3}
\and
  J.~Guibert \inst{4}
\and
  C.~Lobo \inst{5}
}
\offprints{E.~Slezak, slezak@obs-nice.fr}
\institute{
    Observatoire de la C\^ote d'Azur, B.P. 229, F-06304 Nice Cedex 4, France 
\and
  Institut d'Astrophysique de Paris, CNRS, Universit\'e Pierre et Marie Curie, 
  98bis Bd Arago, F-75014 Paris, France 
\and 
    DAEC, Observatoire de Paris, Universit\'e Paris VII, CNRS (UA 173),
    F-92195 Meudon Cedex, France 
\and
    CAI et Observatoire de Paris, 61 Avenue de l'Observatoire, F-75014 Paris,
    France
\and
    Osservatorio Astronomico di Brera, via Brera 28, I-20121 Milano, Italy
}
\date{Received, 1999; accepted,}
\begin{document}

\maketitle

\begin{abstract}
We present two catalogues of galaxies in the direction of the rich
cluster \a4.  The first one includes 3,879 galaxies located in a
region of roughly $\pm$~1.3$^\circ$ from the cluster centre. It has
been obtained from a list of more than 35,000 galaxy candidates
detected by scanning part of a Schmidt photographic plate taken in the
\bj\ band.  Positions are very accurate in this catalogue but
magnitudes are not.  This led us to perform CCD imaging observations
in the V and R bands to calibrate these photographic magnitudes. A
second catalogue gives a list of galaxies with CCD magnitudes in the V
(239 galaxies) and R (610 galaxies) bands for a much smaller region in
the centre of the cluster.

These two catalogues will be combined with a redshift catalogue of 466
galaxies (Durret {\it et al.} 1999) to investigate the cluster properties
at optical wavelengths (Durret {\it et al.} in preparation), as a complement
to previous X-ray studies by a member of our group (Pislar 1998).

\end{abstract}

\section{Introduction}

\a4 is a cluster of richness class 1 (Abell 1958) located at an
average redshift $z=$~0.0331. We performed a detailed analysis of this
cluster from the X-ray point of view, based on ROSAT PSPC data (Pislar
1998).

In the optical, no photometric data had been published, and about 150
redshifts were available from the literature at the beginning of our
study (Beers {\it et al.} 1991, Malumuth {\it et al.} 1992). We undertook
a complete analysis of this cluster, with the aim of obtaining both
photometric and redshift data at optical wavelengths and couple them
with X-ray data. We present here our photometric data. The redshift
catalogue is published in a companion paper (Durret {\it et al.} 1999)
and the analysis of all these combined optical data will be presented in
a coming paper (Durret {\it et al.} in preparation).

\section{The photographic plate data}

\begin{figure}
\centerline{\psfig{figure=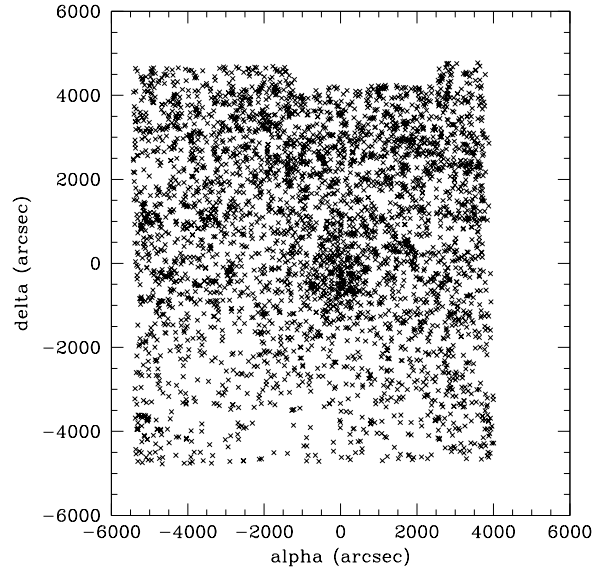,height=8cm}}
\caption[ ]{Spatial distribution of 3,879 galaxy candidates
in the field. }
\protect\label{allplate}
\end{figure}

\begin{figure}
\centerline{\psfig{figure=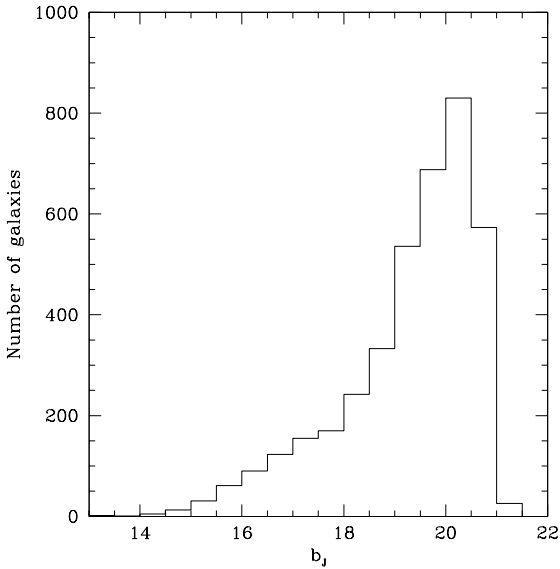,height=8cm}}
\caption[ ]{Magnitude distribution in the $b_{\rm J}$ band of the 
3,879 galaxy candidates in ABCG~496 photographic plate field.}
\protect\label{histobj}
\end{figure}

The photometric catalogue of the galaxies in the direction of the
ABCG~496 cluster of galaxies was obtained by processing the field
number 621 in the SRC-J Schmidt atlas. Part of the glass copy of this
blue plate (IIIaJ$+$GG385) was investigated in June 1993 with the MAMA
(Machine \`a Mesurer pour l'Astronomie) facility located at the Centre
d'Analyse des Images at the Observatoire de Paris and operated by
CNRS/INSU (Institut National des Sciences de l'Univers).  Due to the
cluster location with respect to the plate boundaries, only a
2.5$^{\circ}\times$2.5$^{\circ}$ square region roughly centered on the
cluster coordinates could indeed be searched for objects. At the
cluster redshift, this limited scan is however large enough to secure
a correct investigation of the overall cluster area (radius of
$\sim$~4.5~$h_{50}^{-1}$~Mpc at least).

The algorithms involved in the MAMA on-line mode available at that time
for detecting and measuring astronomical sources are summarized in Slezak
{\it et al.}  (1998). Basically, objects appear to be defined as a set
of connected pixels with intensity higher than typically a 2 sigma threshold
above the local sky background and they are described by their flux (the sum of
background-subtracted pixel values, which lead to what is called hereafter
a ``plate'' magnitude), their area and their elliptical shape parameters.
This approach is efficient in most cases. However, one must expect that this
simple vision model fails for crowded fields where the probability to get
blended objects increases drastically when low detection thresholds are
applied. The galactic latitude of ABCG~496 is $b_{\rm II} \simeq -34.4^{\circ}$
and some blends with stars are indeed included in the 35,541 individual sources
listed in the on-line catalogue we were provided with (cf. the final visual
check of the galaxy candidates which is described below).

The astrometric reduction of the whole catalogue was performed with
respect to 98 stars of the PPM star catalogue (Roeser \& Bastian 1991)
spread over the plate using a 3$^{\rm rd}$-order polynomial
fitting. The residuals of the fit yielding the instrumental constants
were smaller than 0.20 arcsecond and the astrometry of our catalogue
indeed appears to be very good, as confirmed by cross-checking galaxy
coordinates with the literature or the APM database (mean average
separation : 0.7 arcsecond) and by our multi-object fibre spectroscopy
follow-up where the galaxies were always found to be very close to the
expected positions.

The CCD observations needed to calibrate accurately the Schmidt plate
were not available at that time. Hence, a preliminary photometric
calibration of these photographic data has been done using galaxies
with known total blue magnitude. The use of catalogued stars was
rejected since such high-surface brightness objects suffer from too
severe saturation effects (coming both from the emulsion itself and
from the electronic settings of the MAMA facility). So, 40 galaxies
available in the Lyon Extragalactic Database (LEDA, Paturel et
al. 1997) were selected and their magnitudes compared to their
measured blue fluxes providing that no close or overlapping objects
were present as checked from a small scan around each LEDA
galaxy. These 31 undisturbed objects span a 3 magnitude range, but
in a very non uniform way. Hence, the 3$\sigma$ clipping routine
used to compute the best fit further discarded 15 objects. The final
rms on the zero-point is unfortunately not better than 0.4 mag owing
to: i) the quite large uncertainty quoted for the total magnitude
estimate in the database along with their irregular distribution,
ii) the limited magnitude range, and iii) differences in the involved
flux estimates (``plate'' vs. total magnitude).

A basic star-galaxy separation has been performed mainly with respect
to a classical surface brightness criterion. As usual for glass copies
of survey plates, the ability of this criterion to discriminate drops
sharply for objects fainter than approximately 19$^{\rm th}$
magnitude. However completeness and purity of any catalogue are most
of the time competing goals. So, another test based on the elongation
was then performed in order to reject linear plate flaws, as well as
to pick bright elongated galaxies first classified as stars due to
strong saturation effects.  Finally, spurious detections that occur
around very bright stars (area greater than 10$^3$ pixels) due to an
incorrect estimate of the local background were tentatively removed by
checking their location with respect to these bright objects (the
detection processing included no smoothing).  Down to the detection
limit, we obtained a list of more than 4,000 galaxy candidates over
our SRC-J~621 blue field to the detection limit.

The differential luminosity distribution of our catalogue of diffuse
objects indicates that this sample is complete down to the $b_{\rm
J}=$~19.5 magnitude (see Fig.~2). However, its purity is less than the
usual 95\% level for high galactic latitude fields (cf.  Slezak {\it
et al.} 1998). Our spectroscopic run indeed indicates a contamination
level by stars at least three times higher ($\simeq$~20\%, see Durret
{\it et al.} 1999). Such a low success rate may partly be ascribed to
the overall image quality at the plate corner where ABCG~496 is
located. As usual for Schmidt plates the PSF may indeed be quite poor
at the borders, which randomly increases the fuzziness of otherwise
point-like objects and thereby leads to parameter estimates closer to
those of diffuse objects than to genuine star-like ones (cf. Fig.~1
where an overdensity is clearly visible at the edges of the
plate). More generally, one can also question the efficiency of the
classification procedure itself when the number density of stars is
very high. In fact, a selection mostly based on a surface brightness
criterion implies keeping $\simeq$~5\% of the total number of stars in
order to select most of the galaxies. Hence, in most cases, the
contamination level of a quite complete galaxy sample unavoidably
increases with the star number density. The galactic latitude of
ABCG~496 is b$_{\rm II}=-$34.4$^{\circ}$, which is quite close to the
Galactic plane and may explain the high absolute number of
misclassified single stars.  But, for the present data, blended stars
identified as a single diffuse object is the main explanation, as
first noticed during the spectroscopic run and partly expected from
the involved detection software. This is confirmed by a check of our
list of bright ($b_{\rm J}<$~17.5) galaxy candidates against the APM
list for the \#621 SRC-J field. Among our 449 candidates within the
same celestial zone, 389 objects are described as nebular in the APM
catalogue, 52 are classified as star-like, 1 is related to a plate
flaw while 10 have no close counterpart. As evidenced by a visual
check using the DSS, among these 62 discrepant objects, there are 8
genuine galaxies, 2 asymetrical objects, 40 blended stars, 8 single
bright stars (important saturation, diffraction spikes), and 4
star-like objects. So, it appears that only 52 objects are
misclassified (12\% of the total number of candidates), out of which
75\% are merged images, plus two asymetrical objects.

On one hand, this $\simeq$~15\% contamination level is disturbing for studies
involving individual objects picked among galaxy candidates. On the other
hand, its net effect for statistical studies is only a decrease of the contrast
for the signal of interest providing that the misclassified stars are randomly
distributed. So, for such applications, the present photometric catalogue of
ABCG~496 certainly remains valuable.

Table~1 lists the catalogue of galaxy candidates obtained from the
SRC-J~621 plate in the 2.5$^{\circ}\times$2.5$^{\circ}$ field of
ABCG~496. Note that the 101 misclassified stars we were able to
identify during our spectroscopic follow-up have been rejected, as
well as the 54 objects selected by a visual check (the 52
misclassified objects and 2 asymetrical objects described above),
yielding 3,879 entries. The meaning of the columns is the
following~:\\
\noindent
(1) running number;\\
(2) to (4) right ascension (equinox 2000.0);\\
(5) to (7) declination (equinox 2000.0);\\
(8) half-major axis (arcseconds);\\
(9) excentricity $e$ defined as $\sqrt{1-({b\over a})^2}$, where $a$ and 
$b$ are the major and minor axes respectively;\\
(10) position angle of the major axis (from North to East);\\
(11) $b_{\rm J}$ magnitude;\\
(12) and (13) X and Y positions in arcsecond relative to the centre defined
as that of the diffuse X-ray emission of the cluster (see \S 3.1);\\
(14) distance to cluster center in arcseconds;\\
(15) MAMA catalogue reference number.

\section{The CCD data}

\subsection{Description of the observations}

\begin{figure}[ht!]
\centerline{\psfig{figure=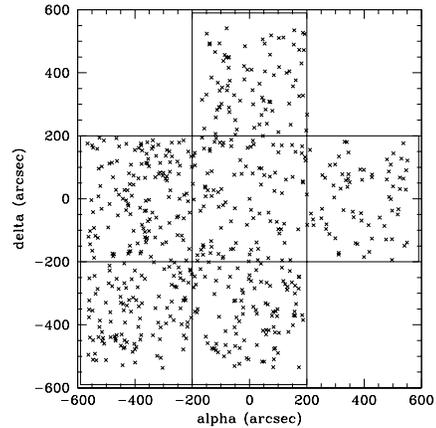,height=6cm}}
\caption[ ]{Positions of the 610 galaxies in our CCD catalogue, with the 
observed fields superimposed. The size of each field is 
6.4$\times$6.4~arcmin$^2$. Positions are drawn relatively to 
the centre defined in the text.}
\protect\label{champsccd}
\end{figure}

The observations were performed with the Danish 1.5m telescope at ESO
La Silla during 2 nights on November 2 and 3, 1994.  A sketch of the
observed fields with the positions of the galaxies appearing in the
CCD catalogue is displayed in Fig.~\ref{champsccd}. The central field
was centered on the coordinates of the cluster center, defined both by
the position of the cD and by the centroid of the X-ray emission
(Pislar 1998): 04$^{\rm h}$33$^{\rm mn}$37.9$^{\rm s}$,
$-13^\circ$15'47"(equinox 2000.0).  There was almost no overlap between
the various fields (only a few arcseconds). Johnson V and R
filters were used. Exposure times were 10~mn for all fields; 1~mn
exposures were also taken for a number of fields with bright stars in
order to avoid saturation. The detector was CCD \#28 with 1024$^2$
pixels of 24~$\mu$m, giving a sampling on the sky of 0.377"/pixel, and
a size of 6.4$\times$6.4~arcmin$^2$ for each field.  The seeing was
poor: 2.4" the first night and 1.1'' the second night.

\subsection{Data reduction}

Corrections for bias and flat-field were performed in the usual way
with the IRAF software. Only flat fields obtained on the sky at
twilight and dawn were used; dome flat fields were discarded because
they showed too much structure.

Each field was reduced separately.  The photometric calibration took
into account a zero point correction, the airmass (AM) and the color index
($V-R$). For the first night, the best fits to obtain the calibrated
magnitudes $V_{cal}$ and $R_{cal}$ from the measured magnitudes $V_m$
and $R_m$ were obtained for the following parameters: \\

$ V_{cal} = V_m - 2.2 - 0.47 \  AM + 0.025 (V-R)_{cal}$,

$ R_{cal} = R_m - 2.07 - 0.51 \  AM + 0.0046 (V-R)_{cal}$, with

$ (V-R)_{cal} = -0.14 + 0.04 \ AM + 1.02 (V-R)_{m} $. \\

\noindent
For the second night, the corresponding relations were: \\

$ V_{cal} = V_m - 2.711 - 0.075 \ AM + 0.009 (V-R)_{cal}$,

$ R_{cal} = R_m - 2.677 - 0.071 \ AM + 0.072 (V-R)_{cal}$, with

$ (V-R)_{cal} = -0.034 - 0.004 \ AM + 0.941 (V-R)_{m} $. \\

The error bars on the various calibration parameters are the
following: V band, first night: $-2.2 \pm 0.1$, $-0.47 \pm 0.08$,
$0.025 \pm 0.025$; R band, first night: $-2.07 \pm 0.15$, $-0.51 \pm
0.11$, $0.0046 \pm 0.037$; V band, second night: $-2.711 \pm 0.004$,
$-0.075 \pm 0.004$, $0.009 \pm 0.003$; R band, second night: $-2.677
\pm 0.002$, $-0.071 \pm 0.002$, $0.072 \pm 0.002$. Note that the color
term for the first night in the R band is in fact undefined; however,
we have kept it for the sake of coherence between all calibrations.
It also appears that both the tranformation and extinction terms for
this first night strongly differ from those for the second night.  For
the first night, the range in standards and in air-mass is smaller,
which may explain the discrepancy, together with perhaps poorer
photometric conditions.

Since the exposure times were the same in V and R, a number of
galaxies were detected in R but not in V. For these objects, a
photometric calibration was performed without using a color term. The 
corresponding relations were:
$ R_{cal} = R_m - 2.06 - 0.51 \  AM $ for the first night and
$ R_{cal} = R_m - 2.70 - 0.03 \ AM $ for the second night.
The corresponding error bars on these parameters are:
$-2.06 \pm 0.14$, $-0.51 \pm 0.11$, $-2.70 \pm 0.04$, $-0.028 \pm 0.028$.

Objects were automatically detected using the DAOPHOT/ DAOFIND tasks
of IRAF. This task first performs a convolution with a Gaussian with
characteristics set according to the seeing in each frame (FWHM of the
star-like profiles in the image) as well as the CCD readout noise and
gain. Objects are then identified as the peaks of the convolved image
which are higher than a given threshold above the local sky background
(chosen as approximately equal to 4 times the {\sl rms} of the mean
sky level on the image). A list of detected objects is thus produced
and interactively corrected on the displayed image so as to discard
spurious objects, add undetected ones (usually close to the CCD edges)
by modifying the detection parameters and dispose of false detections.

We used the package developed by Le F\`evre (Le F\`evre {\it et al.}
1986) to obtain for each field a catalogue with the (x,y) galaxy
positions, isophotal radii, excentricities, major axis, position
angles, and V and R magnitudes within the 26.5 isophote. To perform a
star-galaxy classification based on the compactness parameter
described in Le F\`evre {\it et al.} (1986), we measured the required
information for each object with dedicated software we developed.
Very bright stars are saturated and deviate significantly from the
Gaussian-like PSF involved in the computation of the classification
parameter. They are therefore classified as non stellar objects with
this criterion and had to be eliminated manually. The rms accuracy on
these CCD magnitudes is about 0.1 magnitude, and their errors are in
all cases smaller than 0.2 magnitude.

The astrometry of this CCD catalogue is accurate to about 2.0
arcseconds as verified from the average mutual angular distance
between CCD and MAMA equatorial coordinates for the galaxies included
in both catalogues.

\begin{figure}
\centerline{\psfig{figure=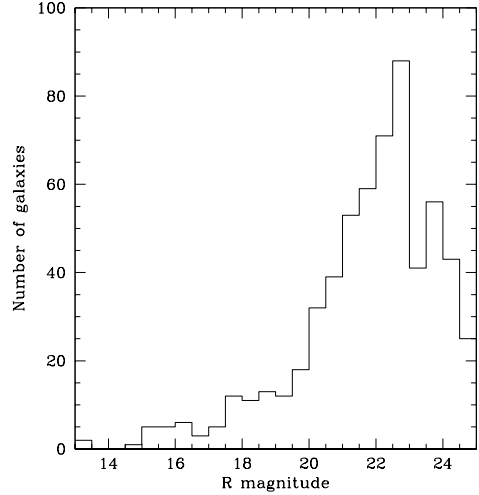,height=7cm}}
\caption[ ]{Histogram of the R magnitudes of the 610 galaxies in the
CCD catalogue.}
\protect\label{ccdrmag}
\end{figure}

The histogram of the R magnitudes in the CCD catalogue is displayed in
Fig.~\ref{ccdrmag}. The turnover value of this histogram is located
around R$\simeq 22.5-23$, suggesting that our catalogue is complete up to
R$\sim 22$.

\begin{figure}
\centerline{\psfig{figure=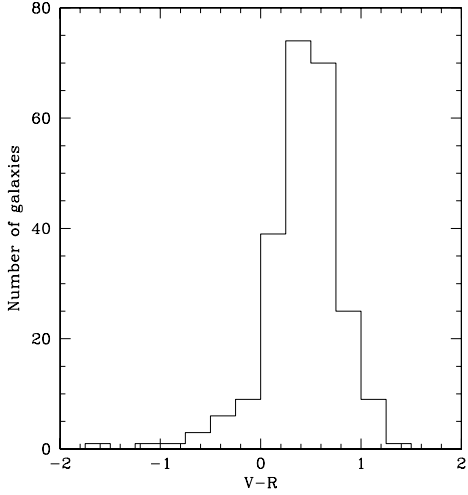,height=7cm}}
\caption[ ]{Distribution of the (V$-$R) colour as a function of R for the 
239 galaxies detected in the V band in our CCD catalogue. }
\protect\label{coul}
\end{figure}

The histogram of the (V$-$R) colour is plotted in Fig.~\ref{coul}. 

\subsection{Transformation laws between the photometric systems}

\begin{figure}
\centerline{\psfig{figure=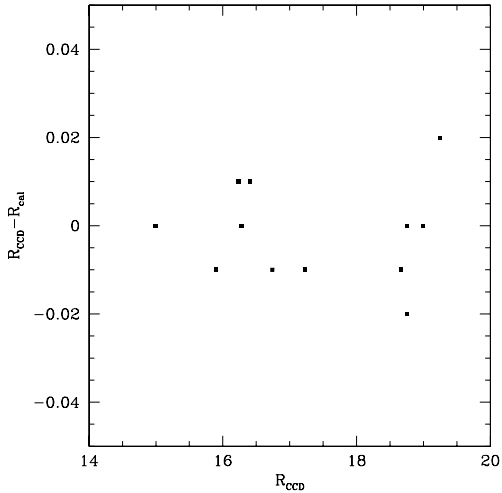,height=7cm}}
\caption[ ]{Difference between the R band CCD magnitude and the R band
magnitude calculated from the photographic \bj\ magnitude (see text) 
as a function of observed R band CCD magnitude. }
\protect\label{rr}
\end{figure}

By identifying galaxies in our CCD catalogue with objects in our
photographic plate catalogue, we derived the following
calibration relations between our photographic plate \bj\ magnitudes and 
our R CCD magnitudes: first night:
$R_{\rm CCD} = b_{\rm J} - 0.28 $, rms=0.05 (5 galaxies);
\noindent
second night:
$R_{\rm CCD} = b_{\rm J} - 0.30 $, rms=0.01 (9 galaxies);
\noindent
mean value:
$R_{\rm CCD} = b_{\rm J} - 0.28 \pm 0.01 $, rms=0.01 (10 galaxies).
We did not include any colour term, because it did not make the fit
any better. The difference between the observed R band CCD magnitude
$R_{\rm CCD}$ and the R magnitude calculated from $b_{\rm J}$ with the
above formula is plotted in Fig.~\ref{rr} as a function of the CCD R
magnitude. This difference is small and does not appear to increase 
with magnitude.

\subsection{The CCD catalogue}

The CCD photometric data for the galaxies in the field of ABCG~496 are given 
in Table~2. The meaning of the columns is the following :\\
(1) Running number; \\
(2) to (4) right ascension (equinox 2000.0); \\
(5) to (7) declination (equinox 2000.0); \\
(8) 26.5 magnitude isophotal radius in arcseconds; \\
(9) excentricity e defined as $\sqrt{1-({b\over{a}})^2}$, where $a$ and 
$b$ are the major and minor axes respectively;\\
(10) position angle of the major axis (in degrees from North to East);\\
(11) and (12) V and R magnitudes;\\
(13) and (14) X and Y positions in arcseconds relative to the X-ray
centre.\\

\section{Summary} 

Our redshift catalogue is submitted jointly in a companion paper
(Durret {\it et al.} 1999). Together with the catalogues presented
here, it will be used to give an interpretation of the optical
properties of \a4 (Durret {\it et al.} in preparation), in relation
with the X-ray properties of this cluster (Pislar 1998).

\acknowledgements {We are very grateful to the MAMA team at
Observatoire de Paris for help when scanning the photographic plate,
and to Cl\'audia Mendes de Oliveira for her cheerful assistance at the
telescope. The help of Fran\c coise Warin in the data reduction is
gratefully acknowledged. We thank the referee, John Huchra, for many
comments which allowed us to improve the paper and in particular for
pointing out mistakes which led us to redo the photometrical
calibration.  C.L. acknowledges financial support by the CNAA (Italia)
fellowship reference D.D. n.37 08/10/1997. }

\end{document}